\RequirePackage{fix-cm}
\documentclass[twocolumn]{svjour3}         
\smartqed  
\usepackage{amsmath,amssymb,amsfonts}
\usepackage{graphicx}
\usepackage{epsfig}
\usepackage{caption}
\usepackage{subcaption}
\usepackage{calc}
\usepackage{bbold}
\usepackage{bbm}
\usepackage{geometry}
\usepackage{adjustbox}
\usepackage{lineno}

\usepackage{color}

 \journalname{Submitted to arxiv.com}
\usepackage[linesnumbered,ruled,vlined]{algorithm2e}

\SetCommentSty{mycommfont}

\SetKwInput{KwInput}{Input}                
\SetKwInput{KwOutput}{Output}              
\begin{document}

\title{Splitter Orderings for Probabilistic Bisimulation
}

\author{Mohammadsadegh Mohagheghi
	  \and
        Khayyam Salehi
}


\institute{M. Mohagheghi \at
              Department of Computer Science, Vali-e-Asr University of Rafsanjan, Rafsanjan, Iran \\
              Tel.: +98-3431312324\\
              \email{mohagheghi@vru.ac.ir}           
           \and
           K. Salehi \at
           Department of Computer Science, Shahrekord University, Shahrekord, Iran\\
           Tel.: +98-3832324401\\
           \email{kh.salehi@sku.ac.ir}
}
\date{}

\maketitle

\begin{abstract}
Model checking has been proposed as a formal verification approach for analyzing computer-based and cyber-physical systems. The state space explosion problem is the main obstacle for applying this approach for sophisticated systems. Bisimulation minimization is a prominent method for reducing the number of states in a labeled transition system and is used to alleviate the challenges of the state space explosion problem.
For systems with stochastic behaviors, probabilistic bisimulation is used to reduce a given model to its minimized equivalent one. In recent years, several techniques have been proposed to reduce the time complexity of the iterative methods for computing probabilistic bisimulation of stochastic systems with nondeterministic behaviors.
In this paper, we propose several techniques to accelerate iterative processes to partition the state space of a given probabilistic model to its bisimulation classes. The first technique applies two ordering heuristics for choosing splitter blocks. The second technique uses hash tables to reduce the running time and the average time complexity of the standard iterative method. The proposed approaches are implemented and run on several conventional case studies and reduce the running time by one order of magnitude on average.

\end{abstract}

\section{Introduction}
\label{intro}
Computers are ubiquitous in modern life, and their failures can have far-reaching implications. Additionally, establishing the correctness of computer systems is a crucial issue since it may jeopardize human life if specific safety systems fail. For example, a miscalculation in launching a rocket can adversely affect the whole project~\cite{clarke2018introduction}.

Testing is a promising technique to assure system correctness. Although it is a common technique, testing cannot cover the whole scenario to verify the correctness of the system~\cite{clarke2018introduction}. On the other hand, some methods utilize mathematical techniques to determine, if the system would operate adequately under all potential circumstances, which are called formal methods. The two most commonly used formal methods are theorem proving and model checking.
The former uses mathematical proof to satisfy the program properties of the system using some possible expert efforts. In contrast, model checking automatically checks that the whole behavior of the system satisfies the desired properties~\cite{baier08}. In this paper, the model checking approach will be considered. 

Model checking is a formal verification approach for analyzing qualitative or quantitative properties of computer systems. In this approach, a Kripke structure or labeled transition system is used to model the underlying system. Also, a temporal logic or automaton is utilized for proposing the required properties. A model checker automatically verifies properties in the proposed model~\cite{baier08,clarke2018handbook}.

Due to some stochastic aspects of many computer systems, it is possible to perform a probabilistic model checking to verify the necessary properties of those systems. Markov decision processes (MDPs) and discrete-time Markov chains (DTMCs) are  extensions of transition systems for modeling stochastic computer systems. DTMCs model fully probabilistic systems, while MDPs model both stochastic and nondeterministic behaviors of computer systems. The probabilistic computational tree logic (PCTL) is used to express properties for verifying MDP and DTMC models~\cite{baier08}.

The main challenge of model checking is the state space explosion problem; that is, by increasing the number of components, the size of models grows exponentially~\cite{baier08,katoen16,parker2003}. This challenge limits the explicit model representation to small ones. Various techniques have been developed in  recent decades to cover this problem. Symbolic model representation~\cite{klein16,parker2003}, compositional verification~\cite{feng2013,forejt2011}, statistical model checking~\cite{agha2018survey,larsen2016statistical}, and reduction techniques~\cite{Hansen2011,Kamal18,kwiatkowska2006symmetry} are the most major techniques that are widely used in model checking tools. Bisimulation minimization is one of the model reduction techniques~\cite{baier20}. This approach can be applied to the verification of security protocols~\cite{abadi1998bisimulation,noroozi2019bisimulation,salehi2022automated}. It defines an equivalence class on the model state space  that can be applied to it. States of each equivalence class are called bisimilar states and satisfy the same set of properties. For a bisimulation relation, the states of any class can be collapsed to one state, and the model will be reduced to a minimized but equivalent one. The quotient model is guaranteed to meet the same set of properties and can be used by a model checker instead of the original~\cite{baier08}.

Depending on the class of transition systems and the underlying properties, several types of bisimulation are defined in the literature. In strong bisimulation, two states, $s$ and $t$, are bisimilar if and only if for each successor state of $s$, there is at least one bisimilar successor state of $t$ and vice versa \cite{groote18}. Two (strongly) bisimilar models satisfy the same set of PCTL formulas~\cite{baier20}. In weak bisimulation, silent transitions are disregarded, and bisimilar states are defined based on a path with some silent moves and a move with the same action~\cite{cattani2002decision,philippou2000weak}. In this paper, we focus on strong bisimulation and propose several heuristic methods in order to reduce the running time of iterative algorithms to compute this kind of bisimulation relation in probabilistic systems. More information about other classes and their algorithms is available in~\cite{baier20,cattani2002decision}.

Several techniques have been proposed to compute probabilistic bisimulation in the literature. The first work to define bisimulation for probabilistic automata and MDPs returns to~\cite{Larsen91,Segala95}. The Definition of strong and weak bisimulation for probabilistic systems with non-determinism and their related algorithms were first proposed in~\cite{stoe2002}. Baier et al. proposed an iterative algorithm for computing probabilistic bisimulation with a time complexity of $O(m\cdot  n\cdot \log(m\cdot n))$, where $n$ is the number of states and $m$ is the number of transitions of the model~\cite{baier2000}. Although several algorithms have been developed for other types of probabilistic systems (such as discrete-time and continuous-time Markov chains or Markov reward models), they rarely consider nondeterministic systems with probabilistic transitions. An efficient algorithm with $O(m\cdot \log(m+n))$ time complexity has been recently proposed in~\cite{groote18}. It is a special case of a more generic partition refinement algorithm, proposed in~\cite{deifel}, that has the $O(m\cdot \log(m+n))$ time complexity. This generic algorithm can be used for a wide class of transition systems, including nondeterministic and probabilistic ones. A scalable method has been proposed in~\cite{castro2020scalable} that defines a bisimulation metric to approximate the bisimulation relation on a given MDP model. Based on the provided metric, any two states that are close to each other are considered bisimilar and are in the same equivalent class. The computed partition may or may not coincide with the exact bisimulation relation. 

To avoid storing the entire state space and transitions of a model, several symbolic bisimulation approaches have been proposed and implemented~\cite{dehnert2013smt,hensel2022probabilistic}. The STORM model checker employs a decision diagram-based data structure as a standard symbolic approach and reports promising results in reducing the running time and consumed memory~\cite{hensel2022probabilistic}. However, symbolic approaches for bisimulation minimization bring several challenges that may influence their performance~\cite{fisler2002bisimulation}. STORM also supports the multi-core bisimulation minimization approach to accelerate the computation of bisimilar partitions~\cite{van2018multi}.

\paragraph{Contributions.} As the main contribution of this paper, two ordering heuristics for selecting splitter blocks are proposed. While the method proposed in~\cite{groote18} uses a random ordering for selecting splitters, our orderings are proposed to reduce the average number that each state is considered in a splitter.
The first heuristic considers the topological ordering of the blocks in the initial partition and uses a queue for inserting each new sub-block. It is optimal for acyclic models, and a heuristic for cyclic ones.
The second heuristic considers block sizes as a criterion for selecting splitters. Smaller blocks have more priority than larger ones and should be selected sooner. A priority queue can be used for selecting the smallest block at each iteration. To avoid the computation overhead of working with the priority queue, we can use several queues instead.
Moreover, a hash table is utilized to avoid the time overhead caused by sorting that is necessary to split the blocks into their sub-blocks. This approach can reduce the time complexity of the current methods. Finally, we experimentally analyze the impact of bisimulation reduction on the overall running time of probabilistic model checking. Although a similar study has been done in the previous works for some other classes of transition systems~\cite{garavel2022equivalence,katoen2007bisimulation,katoen2011ins}, we run these experiments on the class of MDP models and report some positive and negative results concerning these cases.   

%

The structure of the paper reads as follows.
In Section~\ref{sec:preliminary}, some preliminary definitions of MDPs, the PCTL logic, probabilistic bisimulation, and the standard algorithm for computing a probabilistic bisimulation partition are provided.
In Section~\ref{sec:ordering}, the ordering heuristics are presented.
Section~\ref{sec:hash} proposes the approach utilizing hash tables for improving the standard probabilistic bisimulation algorithm.
Section~\ref{sec:experimental} demonstrates the experimental results running on several classes of the standard benchmark models. Finally, Section~\ref{sec:conclusion} concludes the paper and defines some future work.

\section{Preliminaries}
\label{sec:preliminary}
For a finite set \textit{S}, a distribution $\mu$ over \textit{S} is a function $\mu \colon S\rightarrow[0,1]$ such that $\sum_{s\in S}\mu(s) = 1$. We consider \textit{S} as  state space and call every member $s\in S$ a state of \textit{S}. The set of all distributions over $S$ is denoted by $\mathcal{D}(S)$. For any subset $T\subseteq S$ and a distribution $\mu$, the accumulated distribution over $T$ is defined as $\mu[T] = \sum_{s\in T}\mu(s)$.

A partition of \textit{S} is a set $\mathcal{B} = \{ B_i \subseteq S \mid i\in I\}$ of non-empty subsets satisfying $B_i\cap B_j = \emptyset$ for all $i,j\in I$. $(i\neq j)$, and $\cup_{i\in I} B_i = S$. The subset $B_i\in\mathcal{B}$ are called equivalence block. For each partition $\mathcal{B}$ of $S$, an equivalence relation \textit{R} is defined where for each states $s,t \in S$ we have $s  R  t$ \emph{iff} there is a block $B_i\in \mathcal{B}$ where $s,t\in B_i$. For an equivalence relation \textit{R} on \textit{S}, the set of equivalence classes of \textit{R} are denoted by $S / R$. For a state $s\in S$, we define $[s]_R = \{t\in S \mid s R t\}$ and $[s]_{\mathcal{B}} = \{t\in S \mid \exists B_i\in\mathcal{B}, \ \ s,t\in B_i \}$.
For a state $s\in S$, $s/R = \{t\in S \mid: s R t \}$ and for a subset $T\subseteq S$, we define $T/R = \{s\in S \mid \exists t\in T : s R t \}$. A partition $\mathcal{B}_1$ is finer than $\mathcal{B}_2$, or $\mathcal{B}_1$ is a refinement of $\mathcal{B}_2$ \emph{iff} for every block $B\in \mathcal{B}_1$, there is a block $B' \in \mathcal{B}_2$ with $B\subseteq B'$.
One can lift an equivalence relation $R$ on $\mathcal{D}$(S) by defining $\mu R\nu$ \emph{iff} $\mu[C] = \nu[C]$ for every block $C\in S/R$.

\begin{definition} \label{def:mdp}
A Markov decision process (MDP) is a tuple $(S,s_0,Act,\delta,G)$ where:
\begin{itemize}
\item \textit{S} is a finite set of states,
\item $s_0\in S$ is an initial state,
\item $Act$ is a set of finite actions,
\item $\delta: S\times Act \rightarrow \mathcal{D}(S)$ is a (partial) probabilistic transition function, which maps a state of \textit{S} and an action to a distribution of states, and
\item $G\subseteq S$ is the set of goal states.
\end{itemize}
\end{definition}

MDPs are used to model systems with both nondeterministic and stochastic behavior. For each state $s\in S$, $Act(s)$ denotes the set of all enabled actions in $s$. Given a state $s\in S$ in an MDP $M$, an enabled action $a\in Act(s)$ is selected nondeterministically and according to the induced distribution $\mu =\delta(s,a)$ the next state $s'$ is probabilistically selected. We use $|S|$ for the number of states, $|Act|$ for the number of actions, and $|M|$ for the size of the model $M$ which is defined as the total number of its states and the probabilistic transitions.

A $\emph{path}$ in an MDP is a finite or infinite sequence of states and actions the form $(s_i,a_i)$ where $s_i\in S$, $a_i\in Act(s_i)$, $\delta(s_i,a_i)(s_{i+1})>0$ for each $i\geq 0$, and $s_0$ is the initial state of the model. A probabilistic transition is defined as any tuple $(s,a,s')$ where $\delta(s,a)(s') > 0$. To resolve nondeterministic choices of an MDP, the notion of policies (also called adversaries) is used. A (deterministic) \emph{policy} maps each state $s\in S$ to an enabled action $a\in Act(s)$. This mapping may depend on the sequence of state-actions of a path before reaching $s$ in memory dependent policies and only depends on $s$ in memory-less ones.

For any state $s\in S$, we use $Pre(s)$ and $Post(s)$ for the set of predecessor and successor states of $s$ and define them as:

\begin{displaymath}
Post(s) = \{s'\in S | \exists a\in Act(s), \delta(s,a)(s') > 0 \},
\end{displaymath}
\begin{displaymath}
Pre(s) = \{s'\in S | s\in Post(s')\}.
\end{displaymath}
For a subset of states $C\subseteq S$, $pre(C)$ is the set of predecessor states of $\mathit{C}$ and defined as:

\begin{displaymath}
Pre(C) = \cup_{s\in C} Pre(s).
\end{displaymath}

Reachability properties of probabilistic systems are defined as the probability of reaching a set of states of the model. For MDPs, these properties are defined as the extremal (minimal or maximal) probability of reaching a goal state $G$ over all possible policies. In bounded reachabilities, the number of steps that can be taken are limited to a predefined bound.

\begin{definition}[PCTL syntax]\label{def:pctl}
Considering $AP$ as the set of atomic propositions, the syntax of PCTL is as follows:

\begin{displaymath}
\phi ::= true\ |\ c\ |\ \phi\ |\ \phi\land\phi\ |\ \lnot\phi\ |\ \texttt{P}_{\theta}p[\psi]\
\end{displaymath}
\begin{displaymath}
\psi ::= \texttt{X}\phi\ |\ \phi\ \texttt{U}^{\leq k}\phi\ |\ \phi\ \texttt{U}\ \phi
\end{displaymath}
where $c\in AP$, $\theta\in \{<,\leq, > , \geq \}$, $p\in [0,1]$ and $k\in N$. In the PCTL syntax, $\phi$ is a state formula and is evaluated over the states of an MDP model $M$ while $\psi$ is a path formula and is evaluated over the possible paths of the model. State formula are directly used in probabilistic model checking and path formula can only occur inside a state formula as $\texttt{P}_{\theta}p[\psi]$. In the semantic of PCTL, a model state $s\in S$ satisfies the formula $\texttt{P}_{\theta}p[\psi]$ \emph{iff} for each possible policy of $M$ the probability of following a path from $s$ satisfying $\psi$ is in the interval determined by $\theta p$~\cite{Kwia11}. For the path operators $X$ (next), $U$ (until), and $U^{\leq k}$ (bounded until), the semantic of path formula is defined as the standard CTL~\cite{baier08}.
\end{definition}

Formally, PCTL is used to define a set of requirement properties in verification of stochastic systems.  More details about probabilistic model checking of PCTL formula and iterative methods for computing reachability properties are available in~\cite{baier08,forejt2011,katoen16}.

\begin{definition}[Probabilistic Bisimulation]\label{def:prob-bisim}
	For an MDP \textit{M}, an equivalence relation $R\subseteq S\times S$ is a probabilistic (strong) bisimulation for $M$ if and only if for all pairs of states $s,t \in S$, the property $s R t$ implies that for every action $a\in Act(s)$, there is an action $b\in Act(t)$ such that $\delta(s,a)  R  \delta(t,b)$.
\end{definition}
In probabilistic bisimulation, the probability of going to each block is the same for both actions. Two  states $s,t\in S$ are probabilistically bisimilar (denoted by $s\simeq_p t$) if and only if there is a probabilistic bisimulation \textit{R} such that $s R t$. Two MDPs, $M_1$ and $M_2$, are bisimilar if and only if a probabilistic bisimulation relation $R\subseteq S_1\times S_2$ exists between the state spaces $S_1$ of $M_1$ and $S_2$ of $M_2$.

The main characteristic of probabilistic bisimulation is that if two states $s,t\in S$ are bisimilar, then both satisfy the same set of bounded and unbounded PCTL formulas \cite{baier08}. As a result, an MDP \textit{M} can be replaced by a reduced bisimilar one where all bisimilar states of any block $B_i\in \mathcal{B}$ are replaced by one state. Note that in Definition~\ref{def:prob-bisim}, the action names are not important to decide the bisimilarity of two states~\cite{baier08}, and $a$ and $b$ may be two different actions or the same. For probabilistic automata (an extension to MDPs), actions names should be considered in the computation of probabilistic bisimulation~\cite{groote18,stoe2002}.

\subsection{The standard algorithm for computing a probabilistic bisimulation partition}\label{sec:stand_alg}
\begin{figure*}[t!]
	\centering
		\includegraphics[scale=0.45]{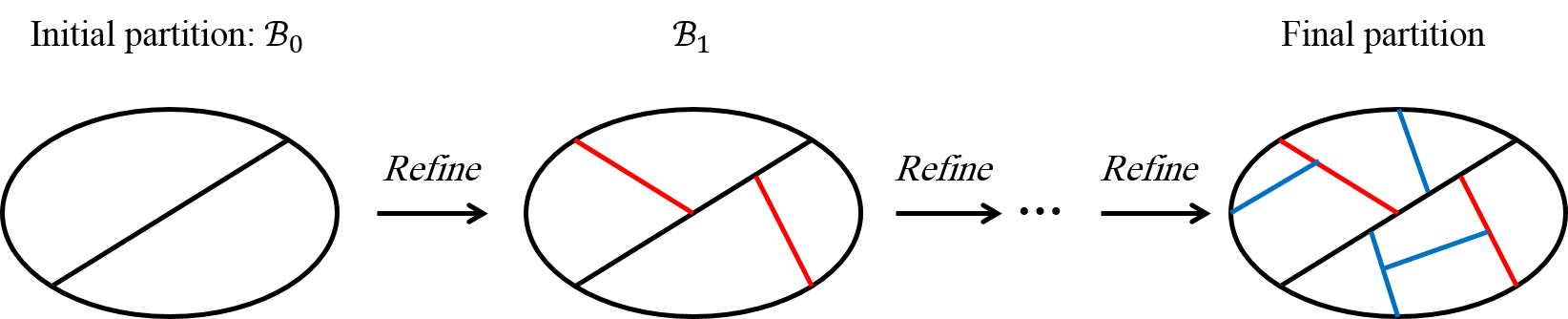}
		\caption{Successive partition refinement procedure}
		\label{fig:refine}
\end{figure*}

Partition refinement is a general algorithm to compute a bisimulation relation for any type of transition system. Starting from an initial partition, the algorithm iteratively refines partitions by splitting some blocks into finer ones. The iterations continue until reaching a fixed point where none of the blocks can be split anymore (Fig.~\ref{fig:refine}).
In each iteration, a splitter block is selected to divide some predecessor blocks into smaller (and finer) ones. The way that the algorithm splits a block depends on the definition of bisimulation for the underlying transition system. Algorithm~\ref{alg:general_bisimulation} presents this approach~\cite{baier08}. This algorithm can be used for any type of transition system by performing appropriate refinement steps for the underlying class of transition system.
%
	\begin{algorithm}[!ht]
	\caption{Partition Refinement}
	\label{alg:general_bisimulation}	
	\KwInput{An MDP model M}
	\KwOutput{bisimulation partition $\mathcal{B}$}
	Initialize $\mathcal{B}$ to a first partition\;
	
	\While{{there is a splitter  for $\mathcal{B}$}}
	{
		Choose a splitter $C$ for $\mathcal{B}$\;
		$\mathcal{B} := \mathit{Refine}(\mathcal{B}, C)$\;
	}
	\Return{ $\mathcal{B}$}\;
\end{algorithm}

In probabilistic bisimulation, the refinement method follows Definition~\ref{def:prob-bisim} and splits the blocks by considering the probability of reaching the splitter $C$. This method is explained in Algorithm~\ref{alg:Partition_Ref}.

\begin{algorithm}[ht]
	\caption{The \textit{Refine} Algorithm}
	\label{alg:Partition_Ref}
	\KwInput{A partition $\mathcal{B}$  and splitter $C$}
	\KwOutput{A refined partition $\mathcal{B}$ according to splitter $C$}
	Initialize $\mathcal{B}$ to a first partition\;
	\ForAll{$B_i\in \mathcal{B}$ s.t. $B_i\cap Pre(C) \neq \phi$}
	{
		\ForAll{$s\in B_i$}
		{
			\ForAll{$a\in Act(s)$}
			{
				Compute $\delta(s,a)[C]$\;
			}
		}
		Define Q = $\{ 0 < q \leq 1\ |\ \exists s\in B_i, \exists a\in Act(s):\delta(s,a)[C] = q\}$\;
		Sort elements of Q\;
		$\textbf{B}_p = \{ B_i \}$\;
		\ForAll{$q\in Q$}
		{
			\ForAll{$B_{i,j}\in \textbf{B}_p$}
			{
				$B' = \{s\in B_{i,j} | \exists a\in Act(s): \delta(s,a)[C] = q \}$\;
				\If{$B'\neq B_{i,j}$ and $B'\neq \phi$}
				{
					Remove $B_{i,j}$ from $\textbf{B}_p$\;
					Add $B'$ and $B_{i,j} \backslash B'$ to $\textbf{B}_p$\;
				}
			}
		}
		Remove $B_i$ from $\mathcal{B}$\;
		Add members of $\textbf{B}_p$ to $\mathcal{B}$\;
	}
	\Return{$\mathcal{B}$}\;
\end{algorithm}

Algorithm~\ref{alg:Partition_Ref} splits any block $B_i$ of the current partition $\mathcal{B}$ into several sub-blocks $B_{i,1}, B_{i,2},\ldots, B_{i,k}$ such that
\begin{enumerate}
	\item $\cup_{1\leq j \leq k}B_{i,j} = B_i$,
	\item $B_{i,j}\cap B_{i,l} = \emptyset$ for $1\leq j < l \leq k$,
	\item for each $1\leq j \leq k$ and every two states $s,t\in B_{i,j}$, it holds that for each action $a\in Act(s)$, there is an action $b\in Act(t)$, where $\delta(s,a)[C] = \delta(t,b)[C]$.
\end{enumerate}

For a given splitter $C$ and every predecessor block $B_i$, the method considers all incoming transitions to the states of $C$ to compute  $\delta(s,a)[C]$, for the related states and actions. After computing these values, the method uses different probability values stored in the set $Q$ for splitting the states of $B_i$ into their sub-blocks. For each probability $q$ and every computed sub-block $B_{i,j}$ of $B_i$, the set of states that can reach $C$ via an action with the probability $q$ is considered in $B'$. Based on this computation, every sub-block $B_{i,j}$ is split into two new ones. Using an optimized data structure, the time complexity of most parts of Algorithm~\ref{alg:Partition_Ref} (except the sorting computation) is linear in the number of incoming transitions to $C$~\cite{groote18}. On the other hand, the time complexity of overall computation for sorting members of the $Q$ sets is in $O(|M| \cdot  \log(| S |))$. More details about this method and its stability conditions are available in~\cite{groote18}.

Algorithm~\ref{alg:general_bisimulation} uses a list of blocks to keep them as a splitter. After refining each block, all computed sub-blocks except the largest one are added to the list to be used as potential splitters. Following this strategy, each state is considered in some splitters for at most $\log(|S|)$ times. As a result, the time complexity of Algorithm~\ref{alg:general_bisimulation} for computing probabilistic bisimulation is in $O(|M|\cdot \log(|S|) + |M|\cdot \log(|Act|))$~\cite{groote18}. For the case of DTMCs where $|S| = |Act|$, this time complexity is expressed as $O(|M|\cdot \log(|S|))$.

Several approaches can be used to compute the first partition $\mathcal{B}\subseteq S\times S$. One can consider $G$ and $S\setminus G$ as the only two blocks of the initial partition and use $G$ as the first splitter. To preserve all reachability properties, the shortest path to $G$ can be used to initialize the $\mathcal{B}$ partition~\cite{Dehn17}. Starting from $G$ and applying a breadth-first search (BFS) in reverse order, all states with the same depth should be put into the same block of the initial partition. Using a finer initial partition, Algorithm~\ref{alg:general_bisimulation} needs a smaller number of iterations to reach the fixed point. Consider the shortest path to compute the initial partition respecting the definition of bisimulation for MDPs and also, the fact that every two bisimilar states, $s_1$ and $s_2$, satisfy the same set of PCTL formulas~\cite{baier08}. Suppose that the shortest paths from $s_1$ and $s_2$ to $G$ (the nearest goal state for each one) are, respectively, $k_1$ and $k_2$ while $k_1 < k_2$. Consider the bounded PCTL formula $\phi =\texttt{P}_{>0} [true\  U^{\leq k_1}\ G]$, where $G$ is the label of goal states. For this formula, we have $\phi\models s_1$, but $\phi \not\models s_2$ because there exists a policy for which the probability of reaching from $s_1$ to $G$ is more than zero, and there is no such policy for $s_2$.

\section{Ordering Heuristics for Choosing Splitters}
\label{sec:ordering}
The probabilistic bisimulation presented in~\cite{groote18} randomly selects blocks of $\mathcal{B}$ as splitters. In the worst case, each state is considered $\log(|S|)$ times in the splitters. In practice, the running time of Algorithm~\ref{alg:general_bisimulation} depends on the order of selecting splitter blocks. Thus, optimal ordering may reduce the running time. In this section, two orderings are proposed to reduce the running time of probabilistic bisimulation.

\subsection{Topological ordering for choosing splitters}

For acyclic models, a topological ordering is defined that can decrease the running time of some standard computations in probabilistic model checking. For example, unbounded reachability probabilities or expected rewards are computed in linear time for an acyclic model if a topological ordering is used for updating state values~\cite{cies08}. The idea of selecting splitters according to an appropriate ordering can reduce the overall number of iterations and the running time of Algorithm~\ref{alg:Partition_Ref}.

\begin{figure*}
     \centering
     \begin{subfigure}{0.45\textwidth}
         \centering
         	\includegraphics[scale=0.4]{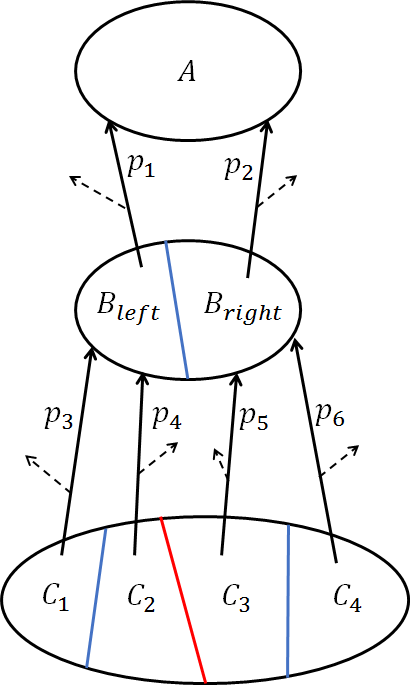}
		\caption{\centering Impact of ordering on the running times}
		\label{fig:splitter-topological-1}
     \end{subfigure}
     \begin{subfigure}{0.45\textwidth}
         \centering
		\includegraphics[scale=0.4]{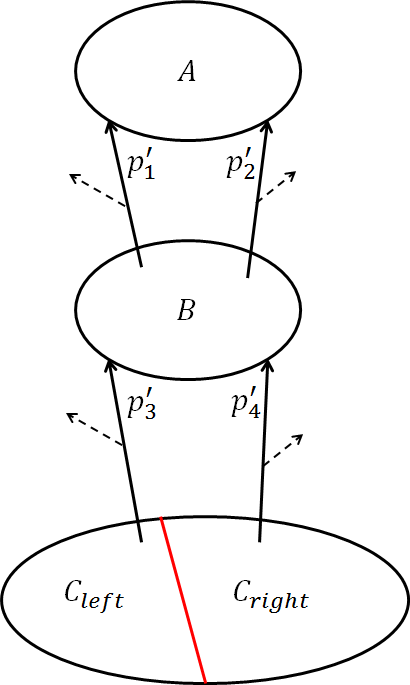}
		\caption{\centering Impact of ordering on the running times}
		\label{fig:splitter-topological-2}
     \end{subfigure}
        \caption{Topological ordering}
        \label{fig:splitter-topological}
\end{figure*}

\begin{example}
	Consider three blocks of a given MDP model during the partition refinement steps, as shown in Fig.~\ref{fig:splitter-topological}. While block $A$ is selected before  block $B$ in Fig.~\ref{fig:splitter-topological-1}, it splits  block $B$ into two new blocks $B_\mathit{left} $ and $B_\mathit{right}$  and considers the smaller one as the next splitter splits $C$ into four blocks. On the other hand, selecting $B$ before $A$ splits $C$ into two new blocks, as demonstrated in Fig.~\ref{fig:splitter-topological-2}. However, $C_\mathit{left}$ and $C_\mathit{right}$ will be split into four blocks $C_1,\ldots,C_4$ after several steps, where $A$ splits $B$ into $B_\mathit{left}$ and $B_\mathit{right}$. In the latter case, the computations for splitting $C$ into $C_\mathit{left}$ and $C_\mathit{right}$ are redundant because the same computations (as in the former case) are needed to split these two blocks into $C_1,\ldots,C_4$. These redundancies can also influence the predecessor blocks of $C$ and bring some other redundancies in the computations of the selected blocks.
\end{example}

For an acyclic model, Algorithm~\ref{alg:topological_order} applies a topological ordering and computes the bisimilar partition with linear time complexity in the size of the model. This algorithm uses a list $L$ of splitters. Each block should be added to $L$ when all of its successor states have been previously used in some splitters. To do so, the incoming transitions to each selected splitter are marked, and a counter can be used to compute the number of unmarked outgoing transitions. Marking all outgoing transitions of a block means that no successor splitter is available for the block and all of its states are bisimilar. The algorithm adds such a block to $L$ to use it as a splitter for its predecessor blocks. The correctness of Algorithm~\ref{alg:topological_order} relies on the fact that there is no cycle among the states of the bisimilar blocks of the model.

	\begin{algorithm}[ht]
	\caption{Topological Partition Refinement}
	\label{alg:topological_order}	
	\KwInput{An MDP model M}
	\KwOutput{A bisimulation partition $\mathcal{B}$}
	Initialize $\mathcal{B}$ to a first partition\;
	L = $\{G\}$\;
	
	\While{L is not empty}
	{
		Remove a splitter $C$ from L\;
		$\mathcal{B} := \mathit{Refine}(\mathcal{B}, C)$\;
		\ForAll{new blocks $D$}
		{
			Mark all transitions between the \textit{D} and \textit{C}\;
			\If{all outgoing transitions from \textit{D} are marked}
			{
				Add \textit{D} to L\;
			}
		}
	}
	\Return{$\mathcal{B}$}\;
\end{algorithm}
For cyclic models, there is no topological ordering~\cite{dai11}. Inspired by the idea of Algorithm~\ref{alg:topological_order}, our first heuristic for choosing splitters is to consider the shortest path to the set of goal states as the initial block ordering and use a queue to keep the new sub-blocks. In each iteration, a splitter \textit{C} is removed from the queue, and if no sub-block of \textit{C} has been previously added to the queue, then Algorithm~\ref{alg:Partition_Ref} uses it as a new splitter. Otherwise, \textit{C} has been split into some sub-blocks by other splitters, and because its sub-blocks are in the queue (except the largest one), the heuristic  disregards \textit{C} for refinement. For any block $B_i\in Pre(C)$, all split sub-blocks $B_{i1},B_{i2},\ldots$, except the largest one, are added to the queue. This heuristic stems from using a splitter before its predecessors' blocks as much as possible.

\subsection{Choosing smallest blocks first}
\begin{figure*}
     \centering
     \begin{subfigure}[b]{0.45\textwidth}
         \centering
		\includegraphics[scale=0.4]{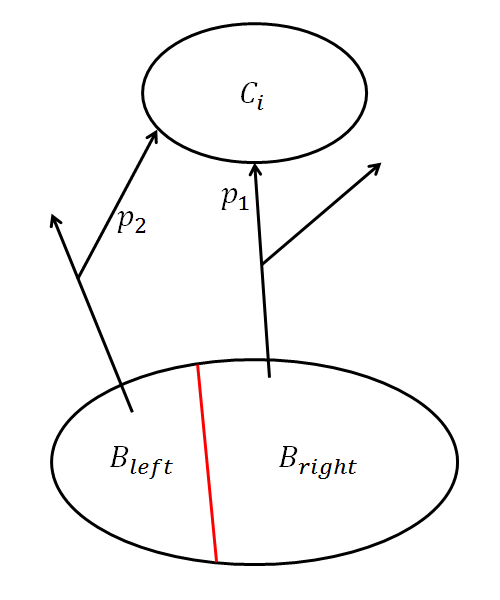}
		\caption{Refinement with large splitters}
		\label{fig:splitter-large}
     \end{subfigure}
     \begin{subfigure}[b]{0.45\textwidth}
         \centering
		\includegraphics[scale=0.4]{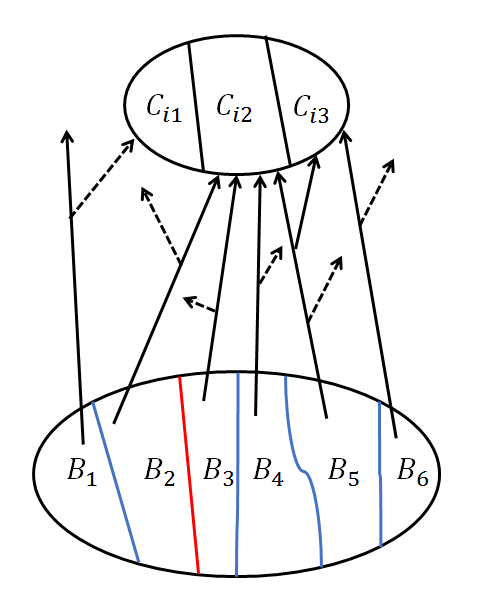}
		\caption{Refinement with small splitters}
		\label{fig:splitter-small}
     \end{subfigure}
        \caption{Size-based splitting}
        \label{fig:splitter-size}
\end{figure*}

In order to minimize the average number of using each state in a splitter,  redundant computations should be avoided. Consider a large block $C_i$ that can be split into several small blocks $C_{i1},C_{i2},\ldots,C_{ik}$. Using $C_i$ as a splitter, Algorithm~\ref{alg:Partition_Ref} may split several predecessor blocks into their sub-blocks. The algorithm will also split these predecessor blocks into some finer sub-blocks when it considers $C_{i1},C_{i2},\ldots,C_{ik}$ as splitters.
The later resulting sub-blocks  are independent of applying $C_i$, and the corresponding computation is redundant; consequently, they can be avoided. To avoid such redundancies, our heuristic method prioritizes smaller blocks to be used as splitters. It can use a priority queue to select the smallest block as the splitter at each iteration. After splitting each block $B_i$, the heuristic removes it from the queue and adds its sub-blocks to the queue.
\begin{example}
	Consider Fig.~\ref{fig:splitter-size}. In Fig.~\ref{fig:splitter-large},  block \textit{B} is split into $B_\mathit{left}$ and $B_\mathit{right}$ by considering $C_i$ as a splitter. In Fig.~\ref{fig:splitter-small}, the sub-blocks $C_{i1}, C_{i2}, C_{i3}$ are used to split  block \textit{B} into sub-blocks $B_1$ to $B_6$. While $B_1$ and $B_2$ are sub-blocks of $B_\mathit{left}$ and $B_3$ to $B_6$ are sub-blocks of $B_\mathit{right}$, the computations of $B_\mathit{left}$ and $B_\mathit{right}$ are redundant and all sub-blocks $B_1$ to $B_6$ can be computed after computing $C_{i1}, C_{i2}, C_{i3}$.
\end{example}
Using \emph{heap} as a standard data structure for a priority queue, each insert and delete operation imposes a $\log(n)$ extra computation that can affect the total running time of the computations. For large blocks, where the number of states is much more than $\log(|S|)$, the running time of using priority queue is negligible, and this data structure can be useful. An alternative approach for small blocks is to use several queues to keep blocks of different sizes. In this approach, we use a queue for the blocks whose size is less than or equal to $\log_2|S|$, and a second queue for those blocks whose size is more than $\log_2|S|$ and less than $c\cdot  \log_2|S|$. The remaining blocks are handled by the priority queue.

\section{Using Hash Tables for Improving Partition Refinement}
\label{sec:hash}
For each splitter \textit{C}, the standard version of Algorithm~\ref{alg:Partition_Ref} (presented in~\cite{groote18}) sorts the members of $Q$ to use its different members for splitting the states of the predecessor blocks. The worst-case time complexity of these computations is in $O(|M|\cdot\log(|Act|))$. To avoid these sortings and alleviate its overhead, our approach is to use a hash table to capture different members of the set \textit{Q}. In this manner, the computed value $\delta(s,a)[C]$ of each splitter \textit{C} is assigned to its associated entry in the hash table. If the entry is empty or its elements differ from $\delta(s,a)[C]$ (a collision happens), this value should be added to the set \textit{Q} and also to the hash table. To have an efficient hash table and minimize its overhead, we consider the four first meaningful digits of $\delta(s,a)[C]$. For non-zero values, a hash function $h$ is defined as:
\begin{displaymath}
	\centering
	h(\delta(s,a)[C]) = \lfloor 10^{4+k}\times \delta(s,a)[C]\rfloor,
\end{displaymath}

where $k$ is the number of immediate zeros after the floating point. As an example, if for some states, actions, and a splitter, we have $\delta(s,a)[C] = .00013760908$, then the hash function computes $h(\delta(s,a)[C]) = 1376$. For the zero value, we define $h(0) = 0$.

Considering the constant time for the computations of a hash table, the total time complexity of this part of computations reduces to $O(|Act|\cdot  \log(|S|))$ because each action is considered at most $\log(|S|)$ times. To keep the keys with collisions, a linked list can be used for each entry; for adding a key to its related entry, it should be compared with all stored keys in the related linked list. Supposing a low frequency of collisions in the hash table, the total time complexity of Algorithm \ref{alg:general_bisimulation} is in $O(|M|\cdot  \log(|S|))$. In the worst case for each splitter \textit{C}, we may have more than $\log(|Act|)$ collisions in the hash table that increase the time complexity of Algorithm~\ref{alg:Partition_Ref}, consequently, increasing the time complexity of Algorithm~\ref{alg:general_bisimulation} to more than $O(|M|\cdot  \log(|Act|))$. However, using a hash table with $|Act|$ entries, the probability of this case, is less than $2^{-\sqrt{(2\cdot |Act|\cdot \log(|Act|))}}$. For this case the hash function is defined as
\begin{displaymath}
	\centering
	h(\delta(s,a)[C]) = \lfloor |Act|\times 10^{k}\times \delta(s,a)[C]\rfloor.
\end{displaymath}

\begin{lemma}\label{lemma1}
	Consider $k$ different keys and a positive integer $t$. For a hash table with $k$ entries and a function $h$ with a uniform distribution of mapping any key to each entry, the probability of having at least $ k\cdot t $ collisions is bounded by $2^{-\sqrt{2\cdot k\cdot t}}$.
\end{lemma}

Using Lemma~\ref{lemma1}, one can control the probability of reaching the worst-case time complexity of Algorithm~\ref{alg:general_bisimulation}. This probability can be reduced by increasing the size of the hash table.

\section{Experimental Results}
\label{sec:experimental}

To show the applicability and scalability of the proposed approaches, we consider nine classes of standard models. These classes include \textit{Coin, Wlan, firewire, Zeroconf, CSMA, mer, brp, leader}, and \textit{Israeli-Jalfon} case studies from the PRISM benchmark suit~\cite{Kwia11}. Except for the \textit{brp} class, which includes DTMC models, the others are MDP ones. More details about these case studies are available in~\cite{baier19,budde20,Kwia11}. To compare our implementation of the proposed heuristics for probabilistic bisimulation with the other tools, we select two state-of-the-art tools, mCRL2~\cite{bunte19} and STORM~\cite{Dehn17}, that provide the most recent approaches \cite{garavel2022equivalence}.

Some information on the selected models,  the experimental results for our implementation,  and the results for the mCRL2 and STORM tools are demonstrated in Table~\ref{table:compare-others}. We implemented the proposed bisimulation algorithm in~\cite{groote18}. The provided information includes the number of states, actions, and transitions of the original case study models and the number of states after applying the bisimulation reduction technique. The experiments have been performed on a machine running Ubuntu 20.04 LTS with Intel(R) Core(TM)i7 CPU Q720@1.6GHz with 8GB of memory. The results include the running time and memory consumption.

We implemented the proposed approaches as an extension of PRISM 4.6, using its sparse engine that is mainly developed in C language. 

It should be noted that the current version of PRISM does not support probabilistic bisimulation for  MDPs. While PRISM and STORM use 8-bytes floating-point representation for storing probabilities, mCRL2 uses fractions of 4-bytes integers. In this way, mCRL2 compensates for the running time and memory consumption for precise computations. On the other hand, STORM follows the proposed algorithm in~\cite{baier2000} with $O(m\cdot n\cdot(\log\ m + \log\ n))$ time complexity. It supports  sparse and BDD-based data structures. For each case study, the lower running times of these two STORM engines are reported. While STORM supports the PRISM modeling language, mCRL2 follows a probabilistic labeled transition system (plts) and considers action labels to distinguish bisimilar states. For a correct comparison between mCRL2 and others, we translated the PRISM plain text models to a corresponding plts, which is executable by the mCRL2 tool. For the initial partition, our implementation considers the sets $G$ and $S\setminus G$ as the first class of blocks. We disable graph-based pre-computations for qualitative reachability analysis in our experiments because different approaches with different impacts on the overall running times are used in the selected tools.

\subsection{Performance analysis for tools and heuristics}
The results of our experiments are demonstrated in Tables~\ref{table:compare-others} and~\ref{table:compare-approaches}. All times are reported in seconds, and memory consumption is in megabytes. In Table~\ref{table:compare-others}, we report some information about the selected models and running the tools on them. The running times of our implementation in PRISM are based on a random splitter ordering, where the splitters are selected randomly. In our implementation and mCRL2, the reported times include the running time for writing the result bisimilar blocks to the files. In most cases, our implementation outperforms both the mCRL2 and STORM tools.

In the \textit{Coin} cases, the running time of our implementation  outperforms mCRL2 by one and STORM by two orders of magnitude. 
In some cases, the mCRL2 process is killed due to the out-of-memory run-time errors. For example, for the case when $n = 5$ and $k = 100$, our implementation in PRISM computes the bisimulation in 6.1 seconds and consumes 1620MB, while STORM computes the same bisimulation in 1178 seconds and consumes 4GB memory, and mCRL2 is killed by out-of-memory error.

In \textit{Zeroconf} models, the running time and memory consumption of our implementation are around 50\% and 20\% of the ones for mCRL2, respectively. In these models, STORM encounters the memory exception.

For all the reported \textit{CSMA} models, our implementation is faster than mCRL2 and STORM. However, for larger values of the parameter \textit{K} (that are not reported here), all tools terminate because of memory limitation. In \textit{firewire} and \textit{Wlan} models, our implementation outperforms STORM by around three orders of magnitude. For large models of these two classes, mCRL2 terminated because of memory limitation, but STORM is able to continue the computations in its BDD-based engine. For \textit{Israeli-Jalfon} models, STORM outperforms mCRL2 in both running-time and memory computations for large models. For the \textit{mer} models, our implantation proposes promising results, while the STORM model checker needs more than one hour for all cases. The mCRL2 tool encounters the segmentation fault error for our given plain-text models.

In most cases, the memory consumption of the proposed implementation in PRISM is less than the other tools. In most cases, except the \textit{brp} ones, STORM reports around 4GB as the peak of memory consumption regardless of the size of the models. For the PRISM and mCRL2 tools, memory consumption depends on the size of the models. Because mCRL2 uses integer value fractions to store probability values, it consumes more memory than the PRISM implementation, and in some cases, it terminates due to memory exceptions while PRISM can compute the bisimulation.

\begin{table*}\centering\small
	\caption{Comparing the performance of computing bisimulation in three tools for the selected MDP and DTMC models.}\label{table:compare-others}
  \begin{adjustbox}{width=\textwidth}

    \begin{tabular}{|c|c|c|c|c|c|c|c|c|c|c|c|c|c|c|c}
		\hline
		Model &  Parameter   & $| S |$ & $|Act|$ & $|Trns|$ & $|S|$ after & \multicolumn{2}{c|}{PRISM} & \multicolumn{2}{c|}{STORM} & \multicolumn{2}{c|}{mCRL2} \\ \cline{7-12}
		Name  &  Val & $\times 10^{-3}$ & $\times 10^{-3}$ & $\times 10^{-3}$ & reduction & time & mem & time & mem & time & mem\\
		\hline
		&K=200&2050& 5534 & 6918 & 91218 & 1.13 &  420MB  &1139 & 3.9GB & 9.02 & 2398MB\\
		Coin & K=300 & 3074 & 8300 & 10374& 136818 & 1.52 & 608MB &2223&4GB&12.8 & 3672MB\\
		(N=4)& K=400& 4098 & 11064 & 13830 & 182418 & 2.07 & 760MB & $>1h$ & 4.1GB & 21 & 4803MB \\
		& K=500& 5122 & 13829 & 17286 & 228018 & 2.43 & 890MB & $>1h$ & 4.2GB & 27.7& 6119MB \\
		\hline
		& K=30 & 2341 & 7832 & 9787 & 33825 & 1.71 &  513MB  & 112 & 3.8GB & 18.9 & 3225MB\\
		Coin & K=50 & 3890 & 13016& 16267& 56325 & 2.98 &  796MB  &  303 & 3.9GB & 31.4 & 5410MB\\
		(N=5)& K=70 & 5439 & 18200& 22747& 78825 & 4.1 &  1127MB  &  554 & 3.9GB & 45.2 & 7202MB\\
		& K=100& 7762 & 25976 & 32467 & 112575 & 6.1 & 1620MB &  1178&  4GB  & - & Killed\\ \hline
		Coin & K=5  & 2936 & 11727& 14635& 12212 & 2.88  &  703MB & 19.9 & 3.4GB&38.9& 4725MB\\
		(N=6)& K=10 & 5731 & 22924& 28632& 24212 & 5.9  &  1329MB &  853 & 309MB & - & killed \\
		\hline
		& K=12 & 3753 & 6898 & 8467 & 1393850 & 8.13  & 895MB & - & Killed & 22.7 & 4020MB\\
		Zeroconf & K=14 & 4426 & 8144 & 9988 & 1666790 & 10.3 & 962MB & - & Killed & 25.9 & 4843MB\\
		(N=1500) & K=16 & 5010 & 9223 & 11307& 1905323 & 11.9 & 987MB & - & Killed & 30.8 & 5626MB\\
		& K=18 & 5476 & 10085& 12359& 2097569 & 13.2 & 1214MB & - & Killed & 37 & 5978MB\\
		& K=20 & 5812 & 10711 & 13124 & 2237272 & 14 & 1530MB & - & Killed & 39.3 & 6522MB\\ \hline
		CSMA	 & K=4  & 1460 & 1471 & 2397 & 23538 & 0.8	& 246MB & 20.3 & 3.9GB &4.5 & 880MB \\
		(N=3)	 & K=5  & 12070& 12108&20215 & 119440 & 7.96 & 1508MB & 143  & 4GB &43.1 & 6834MB\\
		CSMA	 & K=2  & 762  & 826  & 1327 & 9183 & 0.3 & 161MB & 9.2  & 3.6GB &2.5 & 510MB\\
		(N=4)	 & K=3  & 8218 & 8516 & 15385& 45793 & 5.42 & 1197MB & 52.9 & 3.9GB &37.4 & 4311MB\\ \hline
		firewire & ddl=3000& 1634 & 1853 & 1919 & 622127 &  1.24 & 115MB & 1532 & 4GB & 3.43 & 1376MB \\
		(dl=3)   & ddl=10000&5911 & 6711 & 6945 & 2421127 & 5.95 & 924MB & $>1h$ & 4GB & 14.6 & 4813MB \\
		& ddl=15000&8966 & 10181 & 10535& 3706127 & 9.62 & 1715MB & $>1h$ & 4.1GB & - & Killed \\ \hline
		firewire & ddl=3000& 2238 & 3419 & 4059 & 999607 & 2.17 & 439MB & 2283 & 3992 & 6.9 & 2209MB\\
		(dl=36)  & ddl=10000 & 7670 & 11742& 13936 & 3491643 & 9.54& 1323MB & $>1h$ & 4GB & 27 & 6030MB \\
		& ddl=15000&11550& 17687& 20991& 5271643 & 14.8& 2142MB & $>1h$ & 4.1GB & - & Killed \\ \hline
		& m=17 &131&1114& 19497& 4112 & 0.4& 100MB& 4.3 & 3.8GB & 6.6& 730MB\\
		Israeli-  & m=18 &262& 2359 &4129 & 7685  & 1.05& 202MB & 9 & 3825MB& 15.2& 2840MB\\
		Jalfon   & m=19 &524& 4981 &8716 & 14310 & 2.81& 401MB &18.5&3937MB& 34.1& 3288MB\\
		& m=20 &1049&10486&18350& 27012  & 7.31& 798MB&55.6& 3.9GB & 79.2 & 6887MB\\
		& m=21   &2097 & 22020 & 38535  & 50964 & 17.76 & 1620MB & 153 & 4GB & - & Killed\\ \hline
		& n=1000 &5909 & 22688 & 23273& 560048 & 2.34 & 693MB &$>1h$&3.9GB & - & Killed \\
		& n=2000 &11816& 45302 & 46540& 1120048 & 5.01 & 1217MB&$>1h$&3.9GB& - & Killed \\
		mer	  & n=3000 &17723& 68052 & 69807& 1680048 & 8.2 &1812MB&$>1h$&4GB& - & Killed \\
		& n=4000 &23630& 90734 & 93074& 2240048 & 10.73& 2447MB&$>1h$&4GB& - & Killed \\
		& n=5000 &29537&113416 & 116341& 2800048 & 12.98& 2935MB&$>1h$&4.1GB& - & Killed \\ \hline
		& max=150  & 787 & 787 & 1087 & 422554 & 0.44 & 169MB & 2.9 & 210MB & 2.4&606MB \\
		brp  & max=300 & 1567 & 1567& 2167 & 842704 & 1.24 & 292MB & 12.5& 372MB & 5.1& 1342MB\\
		N=400 & max=600 & 3127 & 3127 & 4327& 1683004 & 2.66& 500MB & 51.4 & 691MB & 10.6 & 2732MB \\ \hline
		& max=150  & 1573 & 1573 & 2174 & 844954 & 0.84 & 232MB & 6.2 &  373MB & 5 & 1371MB\\
		brp  & max=300 & 3133 & 3133 & 4334 & 1685104 & 2.13 & 446MB & 26.7& 692MB & 10.7 & 2606MB \\
		N=800& max=600 & 6253 & 6253 & 8654 & 3365404 & 6.17& 954MB & 103 & 1.3GB & 22.5 & 5232MB \\ \hline
		Wlan & ttm=1500 & 3635 & 6351 & 7635 & 35768 & 0.27 & 181MB & 490 & 3.9GB & 8.1 & 3420MB\\
		(N=5) & ttm=3000 & 5989 & 11088 & 12372 & 65768 & 0.54 & 365MB & 1792 & 4GB &  14.4 & 5167MB\\
		& ttm=4500 & 8345 & 15825& 17109 & 95768 &	0.8 & 539MB & $>1h$ & 4.1GB & 20.9 & 7212MB\\ \hline
		Wlan & ttm=1000 & 8093 & 12543& 17668 &	36006 & 0.75 & 290MB & 320 & 3.9GB & - & Killed\\
		(N=6)& ttm=2500 & 12769& 21925& 27051 & 72006 &	1.15 & 355MB & 1900& 4GB & - & Killed\\    \hline
	\end{tabular}
	  \end{adjustbox}

\end{table*}

\begin{table*}\centering
	\caption{Comparing the impact of the proposed heuristics implemented in PRISM on the performance of the partition refinement algorithm for the  Selected MDP classes.}\label{table:compare-approaches}
\begin{adjustbox}{width=\textwidth}
	\begin{tabular}{|c|c|c|c|c|c|c|c|c|c|c|c|}
		\hline
		Model & {\scriptsize Parameter}   & \multicolumn{2}{c|}{random ordering}  & \multicolumn{3}{c|}{topological ordering} & \multicolumn{3}{c|}{size-based ordering}
		& \multicolumn{2}{c|}{hash-table} \\ \cline{3-5}\cline{6-12}
		Name  & {\scriptsize Value} & {\scriptsize running} & {\scriptsize AvgSpl} & {\scriptsize running} & {\scriptsize memory} & {\scriptsize SplAvg} & {\scriptsize running} & {\scriptsize memory} & {\scriptsize SplAvg} & {\scriptsize running} & {\scriptsize SplAvg}   \\
		& & time & & time & overhead & & time & overhead & & time &   \\
		\hline
		&K=200 & 0.42  & 2.43 & 0.38 & 2.3KB & 1.69 & 0.22 & 245KB & 1.02 & 0.22 & 1.02 \\
		Coin & K=300 & 1.32 & 2.48 & 0.98 & 2.3KB & 1.69 & 0.66 & 373KB & 1.02 & 0.66 & 1.02 \\
		(N=4)& K=400 & 1.77 & 2.48 & 1.73 & 2.3KB & 1.69 & 0.9 & 497KB & 1.02 & 0.86 & 1.02 \\
		& K=500      & 2.26 & 2.49 & 1.65 & 2.4KB & 1.69 & 1.14 & 621KB & 1.02 & 1.09 & 1.02 \\
		\hline
		& K=30 & 1.55 & 2.42 & 1.35 & 17.8KB & 1.99 & 1 & 80KB & 1.14 & 0.87 & 1.14\\
		Coin & K=50 & 2.72 & 2.52 & 2.12 & 17.8KB & 1.99 & 1.45 & 132.3KB & 1.14 & 1.3 & 1.14  \\
		(N=5)& K=70 & 3.72 & 2.5 & 2.92 & 17.8KB & 2 & 2.05 & 186.5KB & 1.14 & 1.85 & 1.14\\
		& K=100& 5.53  & 2.51 & 4.19 & 17.8KB & 2 & 2.96 & 266.3KB & 1.14 & 2.62 & 1.14  \\ \hline
		Coin & K=5  & 2.64 & 2.31 & 3.1 & 135.7KB & 2.13 & 1.82 & 40.3KB & 1.13  & 1.57 & 1.14 \\
		(N=6)& K=10 & 5.53 & 2.38 & 6.36 & 135.7KB & 2.17 & 3.73 & 80.2KB & 1.14 & 3.16 & 1.14 \\
		\hline
		& K=12 & 7.7 & 3.09 &  3.15 & 206KB & 2.5 & 1.4 & 9.67MB & 0.7 & 1.32 & 0.72 \\
		Zeroconf & K=14 & 9.8 & 3.1 & 3.9 & 218KB & 2.54 & 1.58 & 11.5MB & 0.73 &  1.49 & 0.73  \\
		(N=1500) & K=16 & 11.8 & 3.14 & 4.5 & 227KB & 2.58 & 1.39 & 13MB & 0.73 & 1.79 & 0.73 \\
		& K=18 & 12.6 & 3.04 & 5.03 & 233KB & 2.6 & 2.11 & 14.1MB & 0.72 & 2.04 & 0.72  \\
		& K=20 & 13.3 & 3.36 & 5.38 & 231KB & 2.6 & 2.26 & 15.7MB & 0.7 & 2.15 & 0.72 \\ \hline
		CSMA	 & K=4  & 0.7  & 2.36 & 0.66 & 274KB & 2.47 & 0.38 & 85.8KB & 1.05 & 0.38 & 1.06 \\
		(N=3)	 & K=5  & 7.06 & 2.55 & 7.44 & 2.48MB & 2.2 & 5.11 & 425KB & 1.26 & 5.87 & 1.26 \\ \hline
		CSMA	 & K=2  & 0.25 & 2.09 & 0.28 & 53.5KB & 2.13 & 0.18 & 31.9KB & 1 & 0.16 & 1 \\
		(N=4) & K=3 & 4.87 & 2.27 & 5.6 & 931KB & 2.66 & 3.22 & 160KB & 1.09 & 3.03 & 1.1 \\ \hline
		firewire & ddl=3K& 3.4 & 2.33 & 2.91 & 511KB & 1.64 & 2 & 2.29MB & 0.65 & 1.87 & 0.65 \\
		(dl=3)   & ddl=10K& 5.4 & 2.33 & 4.89 & 1.88MB & 1.64 & 2.01 & 9.43MB & 0.65 & 1.87 & 0.65 \\
		& ddl=15K& 8.74 & 2.39 & 7.95 & 1.95MB & 1.64 & 3.26 & 14.5MB & 0.63 & 3.01 & 0.64  \\ \hline
		firewire & ddl=3K& 1.97 &2.71 & 1.6 & 2.05MB & 1.88 & 0.64 & 3.52MB & 0.67 & 0.59 & 0.67 \\
		(dl=36)  & ddl=10K& 8.76 & 2.82 & 7.27 & 7.12MB & 1.89 & 2.8 & 12.5MB & 0.66 & 2.62 & 0.66 \\
		& ddl=15K& 13.6 & 2.79 & 11.6 & 10.7MB & 1.89 & 4.43 &  34.7MB & 0.66 & 4.25 & 0.66 \\ \hline
		& m=17 & 0.4 & 2.12 & 0.42& 159.8KB & 2.17 & 0.32& 19.3KB & 0.97 & 0.2 & 0.97 \\
		Israeli-  & m=18 & 1.03& 2.09 & 1.15 & 319KB & 2.24 & 0.79 & 35.8KB & 0.97 & 0.53 & 0.97 \\
		Jalfon   & m=19 & 2.78 & 2.08 & 2.85 & 635.6KB & 2.25 & 2.01& 66.6KB & 0.97 & 1.46 & 0.97 \\
		& m=20 & 7.24 & 2.1 & 7.43& 1.26M & 2.3 & 5.49 & 124.6KB & 0.97 & 3.99 & 0.97  \\
		& m=21& 17.6 & 2.1 & 18.6& 2.5MB & 2.28 & 13 & 234KB & 0.98 & 9.94 & 0.98 \\ \hline
		& n=1000 & 1.92 & 1.39 &1.85& 14.9MB & 1.23 & 0.8 & 1.42MB & 0.96 & 0.81 & 0.96 \\
		& n=2000 & 4.13 & 1.25 & 4.19 & 29.9MB & 1.23 & 1.64 & 2.86MB & 0.96 & 1.68 & 0.96  \\
		mer	  & n=3000 & 6.87 & 1.29 & 6.64 & 44.9M & 1.24 & 2.54 & 4.29MB & 0.96 & 2.53 & 0.96 \\
		& n=4000 & 8.96 & 1.18 & 9.23 & 59.86M & 1.23 & 3.5 & 5.73MB & 0.96 & 3.43 & 0.96 \\
		& n=5000 & 10.79 & 1.33 & 11.78 & 74.84M & 1.23 & 4.31 & 7.16MB & 0.96 & 4.32 & 0.96 \\ \hline
		& max=150  & 0.34 & 2.35 & 0.31 & - & 1.64 & 0.26 & - & 0.64 & 0.34 & 0.64 \\
		brp  & max=300 & 0.98 & 2.33 & 0.61 & - & 1.65 & 0.54 & - & 0.63 & 0.67 & 0.63  \\
		N=400 & max=600 & 2.31 & 2.32 & 1.4 & - & 1.65 & 1.28 & - & 0.63 & 1.29 & 0.63 \\ \hline
		& max=150  & 0.67 & 2.45 & 0.54 & - & 1.65 & 0.48 & - & 0.55 & 0.46 & 0.55 \\
		brp  & max=300 & 1.68 & 2.38 & 1.22 & - & 1.65 & 1.06 & - & 0.55 & 1.05 & 0.55 \\
		N=800& max=600 & 5.46 & 2.36 & 3.5 & - & 1.65  & 2.95 & - & 0.55 & 2.84 & 0.55 \\ \hline
		Wlan & TTM=1500 & 0.09 & 1.03 & 0.09 & 10.9MB & 1.01 & 0.08 & 90.5KB & 0.99 & 0.08 & 0.99  \\
		Wlan & TTM=3000 & 0.14 & 1.03 & 0.15 & 12.4MB & 1.01 & 0.14 & 172KB & 0.99 & 0.15 & 0.99  \\
		(N=5)& TTM=4500 & 0.23  & 1.02 & 0.21 & 15.3MB & 1.01 & 0.21 & 270.5KB & 0.99 & 0.23 & 0.99  \\ \hline
		Wlan & TTM=1000 & 0.22 & 1.01 & 0.17 & 26.5MB & 1 & 0.17 & 75.1KB & 1 & 0.18 & 1 \\
		(N=6)& TTM=2500 & 0.29 & 1.01 & 0.28 & 38.8MB & 1 & 0.26 & 183KB & 1 & 0.33 & 1  \\    \hline
	\end{tabular}
\end{adjustbox}
\end{table*}

In Table~\ref{table:compare-approaches}, we compare the impact of our proposed methods implemented in PRISM on the performance of the probabilistic bisimulation algorithm. We propose the running time and memory overhead of our methods for the selected case study models. The running times only include the computations of bisimulation blocks, excluding the time for writing reduced models to the files and computing quantitative properties. Moreover,  the average number of using each state in a splitter is considered as a criterion to compare the performance of the applied methods. Let \textit{SPLITTERS} be the set of all blocks that are used as a splitter during the partition refinement computations (Algorithm~\ref{alg:Partition_Ref}). We define the average number of using each state in a splitter as $SplAvg$ equals to $(\sum_{C\in SPLITTERS} |C|)/|S|$ and report this value for each approach in the table.

We recall the running times for our implementation of the proposed method in~\cite{groote18} based on a random ordering for selecting  splitter blocks in the column ``Random ordering". For the proposed ordering heuristics, memory overhead includes the maximum extra memory usage for keeping the states in the related lists and priority queues. In our implementation, the hash tables contain 10000 entries, which need at least 40KB of memory to point to the related elements. Because each splitter $C$ is used only once during the computations, our approach needs one hash table for all computations.

In most cases, our proposed methods reduce the running time of the original probabilistic bisimulation algorithm. Experimental results are promising for  size-based ordering, and in most cases, it reduces the running time to half or less compared to the random ordering approach. 

The topological ordering approach reduces the running time of bisimulation minimization for most cases of \textit{Zeroconf}, \textit{firewire}, \textit{brp}, and \textit{Coin} models, although it is not as good as the size-based approach. For these cases, the value of \textit{SplAvg} is reduced when the topological ordering is used for selecting the splitters. For most models of the \textit{CSMA} and \textit{Israeli-Jalfon} cases, the value of \textit{SplAvg} is increased when our topological-based approach is applied. For the \textit{Mer} and \textit{Wlan} cases, the value of \textit{SplAvg} and the running times are near each other for the random and topological ordering approaches. While for most cases, \textit{SplAcg} is more than two, this value is near one for the \textit{Mer} and \textit{Wlan} cases. For these two classes, a large part of states are either in $G$, or cannot reach this set. Hence, a small part of $S$ remains for the iterative partition refinement computations. The size-based ordering method reduces the value of \textit{SplAvg} to around one or less in most cases. On the other hand, this value is independent of the size of MDPs among the same models of most classes. The only exception is for the \textit{CSMA} models. As a result, the running time of the bisimulation method with size-based ordering is near linear in the size of models for the selected benchmark sets.

The proposed technique in Section~\ref{sec:hash} for using the hash table to improve the time complexity of the bisimulation algorithm proposes a slight improvement in practice in the performance of the bisimulation algorithm. Note that the main benefit of using a hash table is to reduce the probability of reaching the worst-case time complexity of Algorithm~\ref{alg:general_bisimulation} for MDP models, as is described in Lemma 1. However, this method may work faster or slower than the others in practice. In all cases,  memory overheads are less than 100MB and less than 5\% of the memory consumption of the original implementation, which shows that the proposed heuristics are completely feasible. While the size-based ordering approach is faster than the topological ordering, for some cases,  its memory overhead is more than memory overhead of the topological ordering approach, and for other cases, it is less. The main reason that the memory overhead of these approaches is different from one class to the other is that it depends on the structure of the model and the maximum number of splitters that are kept in the related list or priority queue. In all cases, the memory overhead of applying hash table is less than 50KB because of rare collisions in its entries in our experiments.

\subsection{Impact of bisimulation on the overall running time of probabilistic model checking}
\begin{figure}
	\begin{center}
		\includegraphics[scale=0.5]{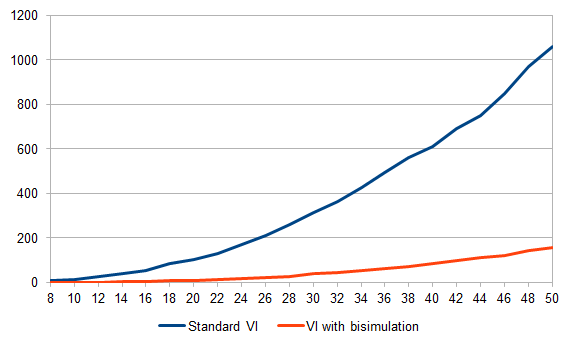}
		\caption{Standard model checking vs. applying bisimulation for the \textit{Coin} models in second}
		\label{fig:coin-diagram}
	\end{center}
\end{figure}

\begin{figure}
	\begin{center}
		\includegraphics[scale=0.5]{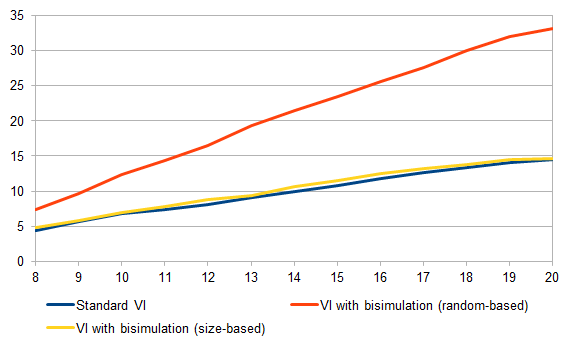}
		\caption{Standard model checking vs. applying bisimulation for the \textit{Zeroconf} models in second}
		\label{fig:zer-diagram}
	\end{center}
\end{figure}

\begin{figure}
	\begin{center}
		\includegraphics[scale=0.5]{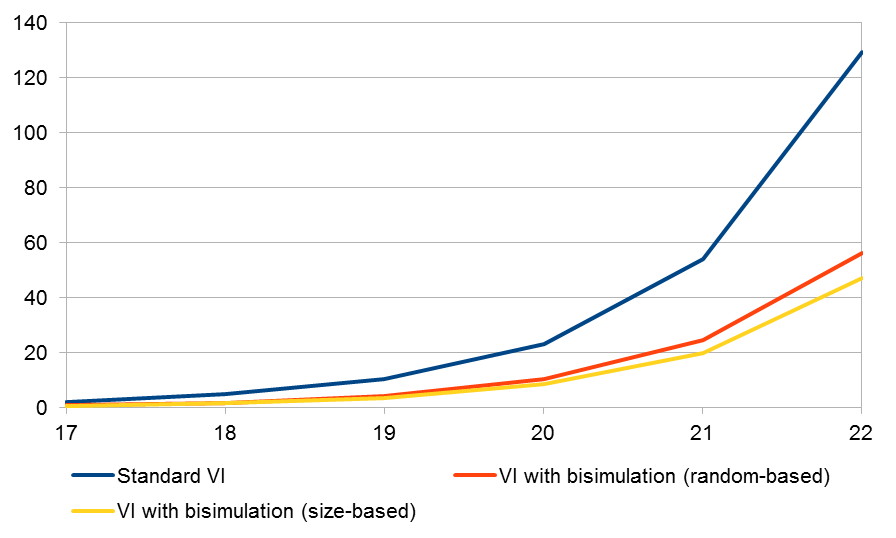}
		\caption{Standard model checking vs. applying bisimulation for the \textit{Israeli-Jalfon} models in second}
		\label{fig:ij-diagram}
	\end{center}
\end{figure}

\begin{figure}
	\begin{center}
		\includegraphics[scale=0.5]{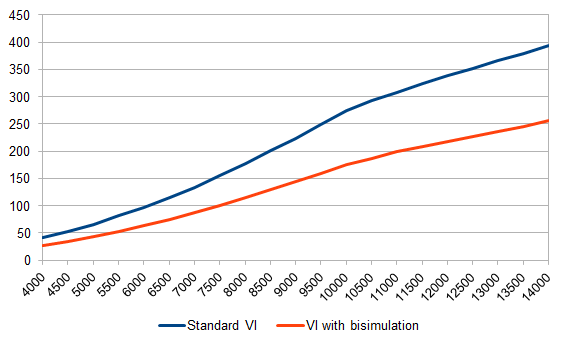}
		\caption{Standard model checking vs. applying bisimulation for the \textit{firewire} models in second}
		\label{fig:fire-diagram}
	\end{center}
\end{figure}

\begin{figure}
	\begin{center}
		\includegraphics[scale=0.5]{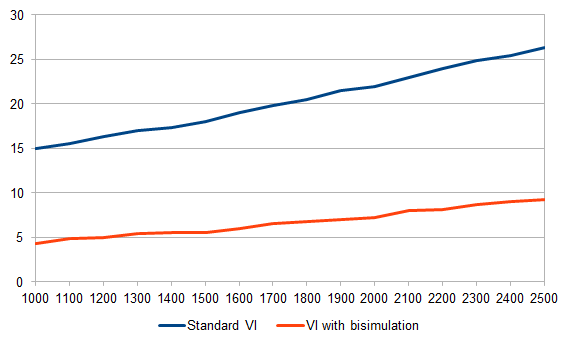}
		\caption{Standard model checking vs. applying bisimulation for the \textit{Wlan} models in second}
		\label{fig:wlan-diagram}
	\end{center}
\end{figure}

\begin{figure}
	\begin{center}
		\includegraphics[scale=0.5]{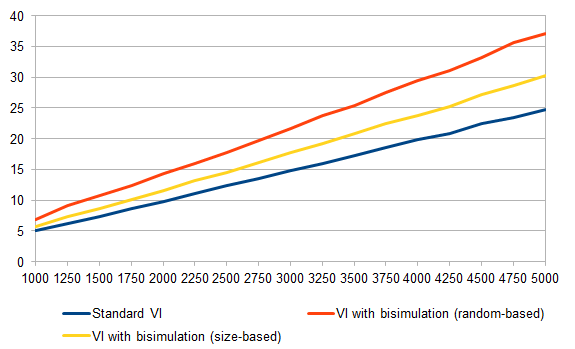}
		\caption{Standard model checking vs. applying bisimulation for the \textit{MER} models in second}
		\label{fig:mer-diagram}
	\end{center}
\end{figure}

\begin{figure}
	\begin{center}
		\includegraphics[scale=0.5]{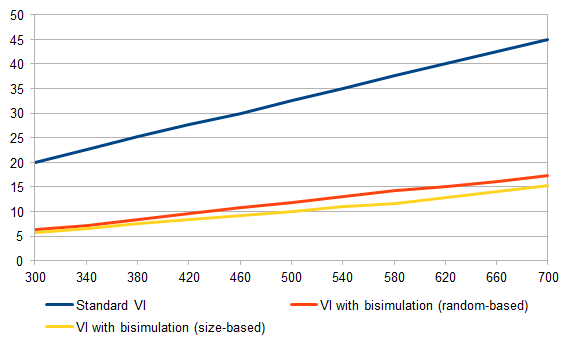}
		\caption{Standard model checking vs. applying bisimulation for the \textit{brp} models in second}
		\label{fig:brp-diagram}
	\end{center}
\end{figure}

To study the impact of bisimulation reduction on the running time of probabilistic model checking, we consider seven classes of case study models, including \textit{Coin}, \textit{Zeroconf}, \textit{Israeli-Jalfon}, \textit{firewire}, \textit{Wlan}, \textit{mer}, and \textit{brp}. For each class, we consider a set of models by setting different values to their parameters. For models of each class, we follow three approaches: (1) running the standard probabilistic model checking without bisimulation, (2) running with the random ordering bisimulation, and (3) running with size-based bisimulation. For an iterative computation method to approximate reachability probabilities, the Gauss--Seidel version of the value iteration method is used. The results of these experiments are demonstrated in Figs.~\ref{fig:coin-diagram}--\ref{fig:brp-diagram}. In each figure, the vertical axis shows the parameter value for the models, and the horizontal axis determines the running time in seconds. In \textit{Coin} and \textit{Zeroconf} cases, \textit{K} is considered as the parameter to have different models. In \textit{firewire} and \textit{Wlan} cases, we, respectively, consider \textit{deadline} and \textit{Trans\_Time\_Max} as the parameter. In the \textit{mer} and \textit{Israeli-Jalfon} cases, the parameter \textit{n} is considered.

The best results are for the class of \textit{Coin} models in Fig.~\ref{fig:coin-diagram}. In this case, the running time of computing bisimulation is negligible compared to the running time of the other numerical computations, and applying this minimization approach reduces the overall running times by one order of magnitude. The number of states of these models after applying bisimulation minimization is less than 10\% of the number of states of the original models, as reported in Table~\ref{table:compare-approaches}. 

In the case of \textit{Zeroconf} (see Fig.~\ref{fig:zer-diagram}), there is a meaningful difference among the applied bisimulation heuristics. Using bisimulation with the random splitter ordering, the computation overhead is more than its benefit, and the overall running times are increased. In this class, iterative computations for reachability probabilities converge fast. Using the proposed size-based heuristic improves the performance such that the running time of applying standard value iteration on the original \textit{Zeroconf} models is near the overall running time of computing bisimulation and applying value iteration on the reduced models. For this class, the size of the reduced models is around 40\% of the size of the original ones, which prohibits bisimulation from reducing the overall running time. 

Fig.~\ref{fig:ij-diagram} for the \textit{Israeli-Jalfon} class shows that the running times are reduced to around half when the bisimulation minimization is applied. In this case, the bisimulation method reduces the size of models to less than 10\% of the original models. Size-based bisimulation reduces around 25\% of the overall running times comparing to the random ordering approach. In Fig.~\ref{fig:fire-diagram} for the set of \textit{firewire} models, bisimulation reduces the overall running times by 30\%. Because the running times of the iterative computations are high, there is no significant difference in the overall times when using different splitter ordering approaches. 
The results of our experiments for the \textit{Wlan} class of models are proposed in Fig.\ref{fig:wlan-diagram}. Although bisimulation results in a significant model reduction for these cases, the overall running times are reduced to less than 40\% after applying bisimulation because a main part of the state is in goal states $G$. 

The results for the \textit{Mer} cases show that the overall running time is increased when the bisimulation minimization methods are applied (Fig.~\ref{fig:mer-diagram}). For this class,
more than 60\% of states are in $G$ and  disregarded for the iterative computations. 

In Fig.~\ref{fig:brp-diagram} for the \textit{brp} DTMC models, applying bisimulation reduces  the running times to less than 30\% of the running time of the standard iterative method without using bisimulation. Although the number of states of the reduced models is around 50\% of the number of states of the original ones, applying bisimulation reduces the number of iterations for computing reachability probabilities.

\section{Conclusion}
\label{sec:conclusion}
In this paper, two approaches to improve the performance of the standard algorithms for computing probabilistic bisimulation in MDPs were proposed.
In the first approach, two heuristics, including topological and size-based ordering, were proposed to determine the ordering of splitters. The second approach used hash tables to reduce the number of comparisons for splitting the blocks of states. Experimental results demonstrated that, in most cases, our approaches outperform the previous algorithms and the other state-of-the-art tools. The impact of bisimulation minimization on the running time of probabilistic model checking was  reported as well. For future work, the applicability of the proposed techniques can be studied on other classes of transition systems, such as probabilistic automata or continuous-time Markov chains. Also, we aim to apply these approaches to other fields, for example, protocol security.


%
\section*{Conflict of interest}
 The authors declare that they have no conflict of interest.


\end{document}